# End-to-end simulations of the Visible Tunable Filter for the Daniel K. Inouye Solar Telescope

Wolfgang Schmidt, Matthias Schubert, Monika Ellwarth, Jörg Baumgartner, Alexander Bell, Andreas Fischer, Clemens Halbgewachs, Frank Heidecke, Thomas Kentischer, Oskar von der Lühe, Thomas Scheiffelen, Michael Sigwarth

Kiepenheuer-Institut für Sonnenphysik, Schöneckstraße 6, 79104 Freiburg, Germany

## ABSTRACT

The Visible Tunable Filter (VTF) is a narrowband tunable filter system for imaging spectroscopy and spectropolarimetry based. The instrument will be one of the first-light instruments of the Daniel K. Inouye Solar Telescope that is currently under construction on Maui (Hawaii). The VTF is being developed by the Kiepenheuer Institut fuer Sonnenphysik in Freiburg as a German contribution to the DKIST.

We perform end-to-end simulations of spectropolarimetric observations with the VTF to verify the science requirements of the instrument. The instrument is simulated with two Etalons, and with a single Etalon.

The clear aperture of the Etalons is 250 mm, corresponding to a field of view with a diameter of 60 arcsec in the sky (42,000 km on the Sun). To model the large-scale figure errors we employ low-order Zernike polynomials (power and spherical aberration) with amplitudes of 2.5 nm RMS. We use an ideal polarization modulator with equal modulation coefficients of $3^{-1/2}$ for the polarization modulation

We synthesize Stokes profiles of two iron lines (630.15 nm and 630.25 nm) and for the 854.2 nm line of calcium, for a range of magnetic field values and for several inclination angles. We estimated the photon noise on the basis of the DKIST and VTF transmission values, the atmospheric transmission and the spectral flux from the Sun.

For the Fe 630.25 nm line, we obtain a sensitivity of 20 G for the longitudinal component and for 150 G for the transverse component, in agreement with the science requirements for the VTF.

## 1. INTRODUCTION

### 1.1 The Daniel K. Inouye Solar Telescope (DKIST)

The DKIST with its 4 m aperture offers unique capabilities for groundbreaking observations and quantitative analysis of photospheric and chromospheric magnetic activity on a variety of spatial scales. The significantly increased resolution and the large photon collecting area of the DKIST compared to its predecessors open up new discovery space. One of the most challenging goals of the DKIST is to perform highly precise measurements of processes in the outer solar atmosphere on their intrinsic physical scales of a few tens of km. Thanks to a site with low scattered light background, the DKIST is very well suited for chromospheric and coronal spectroscopy. The telescope is presently under construction at the Haleakala Observatory on Maui (Hawaii)[1].

### 1.2 DKIST first light instruments

The DKIST will start science operations with a suite of first-light instruments, including the Visible Broad-band Imager (VBI[2]), the Visible Spectro-Polarimeter (ViSP[3]), the Cryogenic Near-Infrared Spectro-Polarimeter (Cryo-NIRSP[4]), the

Diffraction-Limited Near-Infrared Spectro-Polarimeter (DL-NIRSP5), and the Visible Tunable Filter (VTF). The Kiepenheuer-Institut für Sonnenphysik (KIS) is developing the VTF as a German contribution to the DKIST.

**1.3 The Visible Tunable Filter**

The VTF is an imaging spectro-polarimeter based on tunable Fabry-Perot Etalons. Filter-based instruments have important advantages over grating spectrometers, since they provide spectroscopic or spectro-polarimetric information on an extended field of view within seconds at very high spatial resolution. This is especially important for the characterization of short-lived phenomena with life times of the order of one minute or less. The quality of the spectral line profiles may be compromised by variable seeing conditions during observations. The VTF is designed for diffraction-limited imaging in the visible and near infrared, at high temporal and medium spectral resolution.

## 2. PHOTON FLUX AT VTF DETECTORS

**2.1 Instrument transmission**

For the VTF, protected Ag coatings with a reflectivity of 97.5%were assumed for all mirrors, lenses with 0.5% AR coatings, two Fabry-Perot etalons with transmissions of 0.9 each, and a prefilter with a peak transmission of 80%. The total transmission of the VTF is shown in Fig. 1 for Doppler imaging or intensity imaging mode. When the VTF is used in polarimetric imaging mode, a polarizing beam splitter divides the modulated beam 50:50 and sends it to two cameras.

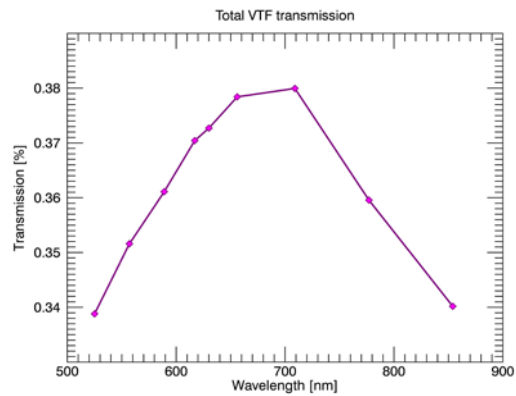

Figure 1. Transmission of the VTF in Doppler and intensity imaging mode.

**2.2 Photon flux**

The photon flux in the detector plane of the VTF was calculated for wavelength bands between 525 nm and 854 nm. The solar spectral flux at disk center was taken from the Neckel-Labs solar flux atlas[6], for the transmission of the Earth atmosphere above Haleakala we used the numbers given in Allen' Astrophysical quantities[7]. Here we included the transmission of DKIST telescope and the VTF proper, as a function of wavelength, and the wavelength-dependent spectral resolution of the VTF. The specified quantum efficiency of the focal-plane detectors was used to estimate the number of detected photons. Figure 2 shows the number of detected photons on each detector in the polarimetric channel of the VTF for a single 25 ms exposure for selected wavelength bands that contain scientifically interesting spectral lines. The numbers refer to the spectral continuum. The peak flux is shifted to the red, compared to the solar spectral flux. This is due to the increase of the spectral resolution element of the VTF proportional to the square of the wavelength. For a single exposure, the SNR at 630 nm is about 200.

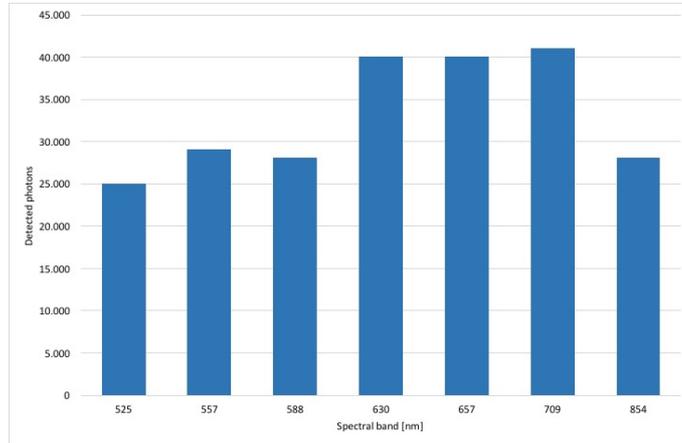

Figure 2. Detected photons per detector for the VTF in polarimetric imaging mode, for 25 ms exposure time.

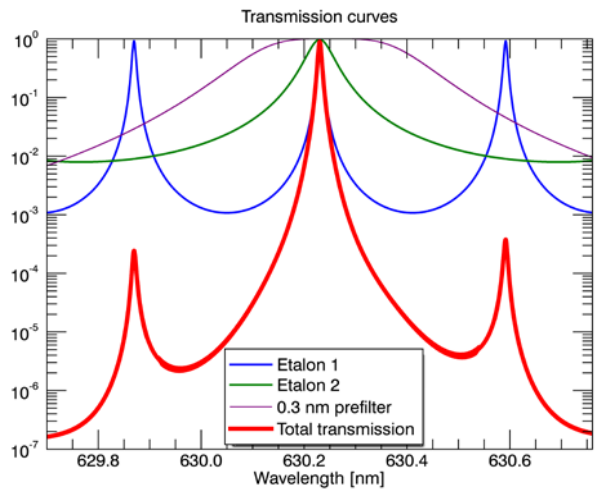

Figure 3: Transmission profile of the VTF in a 1 nm wavelength interval around 630 nm, with 2 Etalons and a 0.3 nm prefilter. The transmission profiles of the Etalons are shown the blue and green, the prefilter profile (2-cavity filter) is depicted in purple. The red line is the total transmission profile.

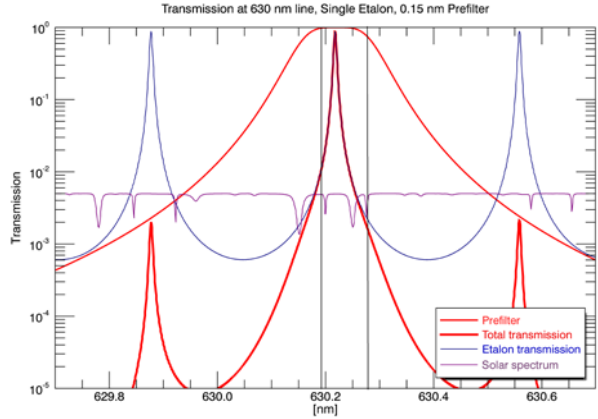

Figure 4: Transmission profile of the VTF in a 1 nm wavelength interval around 630 nm, for a single Etalon and a 0.15 nm prefilter. The solar spectrum is indicated in purple. The transmission profile of the Etalon is shown the blue, the thin red line is the prefilter profile (2-cavity filter). The thick red line is the total transmission profile.

## 3. VTF INSTRUMENT SIMULATOR

### 3.1 Treatment of the Etalons

For the simulations we considered the telecentric configuration of the etalons in an f:200 beam, and their placement at a certain distance from the focal plane. We also included cavity errors of the etalons of 2.5 nm RMS, and a micro-roughness of the etalon plates of 0.1 nm RMS. These values were measured on a full-size prototype plate in the course of the development of the VTF etalons[8]. Thanks to the large f-ratio, we could neglect the effect of pupil apodization[9]. The polarization properties of the etalons were also neglected, since the incidence angle of the light beam deviates only very little from the normal direction[10]. A detailed description of the instrument simulator can be found in the Dissertation of M. Schubert[11].

### 3.2 Interference filters

Narrowband interference filters are employed to suppress all unwanted transmission bands of the Fabry-Perot Etalon. The free spectral range (FSR) of the Etalons depend strongly on wavelength, and this determines the required filter width. We used 2-cavity narrowband interference filters to suppress the unwanted transmission maxima of the Etalons. The filter curves are modeled with the relation

$$T = T_0 \cdot \left[1 + \left(\frac{2\Delta\lambda}{w}\right)^{2N}\right]^{-1},$$

where $T_0$ is the filter transmission at the central wavelength of the filter, w, is the full width at half maximum, and $\Delta\lambda$ is the distance from the central wavelength. The parameter N denotes the number of cavities of the filter coating. Narrowband filters typically have 2 or 3 cavities. Additional cavities improve the performance of the filter bandpass by producing steep transmission profiles with nearly rectangular shape. At the same time, out-of-band blocking increases. Figures 3 and 4 show examples of bandpass filters with half-widths of 0.30 nm and 0.15 nm, respectively.

We focused our simulations on the polarimetric imaging mode of the VTF, since this is the default mode of operation, and it is also the most demanding one. We used an ideal 4-state polarization modulator[12] that is described by the following equation:

$$\begin{pmatrix} I_1 \\ I_2 \\ I_3 \\ I_4 \end{pmatrix} = \begin{pmatrix} 1 & \frac{1}{\sqrt{3}} & \frac{1}{\sqrt{3}} & \frac{1}{\sqrt{3}} \\ 1 & \frac{1}{\sqrt{3}} & -\frac{1}{\sqrt{3}} & -\frac{1}{\sqrt{3}} \\ 1 & -\frac{1}{\sqrt{3}} & -\frac{1}{\sqrt{3}} & \frac{1}{\sqrt{3}} \\ 1 & -\frac{1}{\sqrt{3}} & \frac{1}{\sqrt{3}} & -\frac{1}{\sqrt{3}} \end{pmatrix} \cdot \begin{pmatrix} I \\ Q \\ U \\ V \end{pmatrix}$$

The column vector on the right hand side denotes the Stokes parameters[13], where I is the total intensity of polarized and unpolarized light, Q is the intensity difference between horizontally and vertically linear polarized light, Q is the same, but for $+45^0$ and $-45^0$, and V is the intensity difference of left and right circular polarized light. The equation is used to produce the modulated intensities $I_1$ to $I_4$ from the computed synthetic Stokes profiles (see Sect. 3.3). In addition, we carried out simulations for the Doppler imaging mode.

### 3.3 Line synthesis

Stokes profiles $S = (I \quad Q \quad U \quad V)^T$ of the neutral iron lines at 630.15 nm and 630.25 nm were synthesized with the SIR code[14] for a range of magnetic field values between 10 and 1000 G, and for inclination angles, between 0 and 90 degrees. The computed wavelength range was 1 nm, centered at 630.2 nm, including all solar lines in that wavelength interval, and the step width was 0.1 pm. Figure **5** shows the Stokes I and V profiles of an example, for a magnetic field strength of 100 G and an inclination angle (w.r.t. vertical) of 10 degrees. The Ca 852.2 nm line was synthesized with the Non-LTE code NICOLE[15], also for a wavelength interval of 1 nm. In both cases, the Harvard Smithsonian model[16] of the solar atmosphere was used.

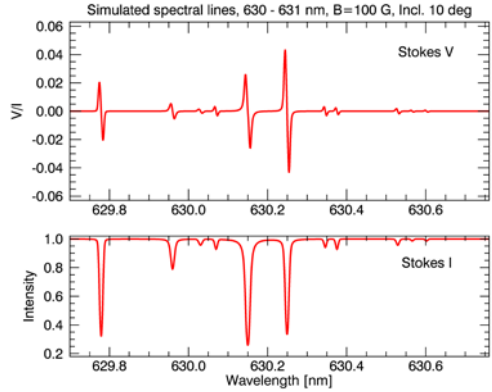

Figure 5: Synthetic spectral lines in the range 629.7 nm to 630.7 nm. The SIR code was used, together with an HRSA standard atmospheric model. The magnetic field strength is 100 G and the inclination is 10°.

### 4. ETALON CAVITY ERRORS

For the very large-scale Etalons needed for the VTF, having a constant air gap across the clear aperture with a diameter of 250 mm is one of the great challenges. Deviations from a perfect air gap result from figure errors of the two Etalon plates. These figure errors may have different causes: (i) residual polishing errors, (ii) gravity, (iii) thickness variation of the coatings, coating stress, or a combination thereof.

We distinguish between small-scale variations (< 1 mm) "micro-roughness" and large-scale variations (>> 1 mm) "figure errors", measured across the Etalon surface. Since the Etalons are mounted telecentrically, i.e. in an intermediate focal plane, the effects of such variations are imaged to the detectors in the focal plane. Therefore, we show the results of this study projected onto the focal plane of the VTF. The clear aperture of the Etalons of 250 mmm corresponds to a field of view of 60 arcsec in the sky, i.e., a micro-roughness of 0.1 mm on the Etalon corresponds to 0.024 arcsec. The air-gap variations are ultimately caused by figure errors of the two high-reflectivity coated inner surfaces of the two Etalon glass plates.

From the coating studies[8] for the VTF Etalons we know that the micro-roughness is of the order of 0.1 nm. This very small value is included in the simulations, but the effect on the results is negligible. In this paper, we therefore concentrate on the effects of large-scale figure errors on the instrument performance.

### 4.1 Figure errors

To investigate the effect of large-scale variations of the air gaps of both VTF Etalons we perform end-to-end simulations of the VTF using synthetic spectral lines derived from numerical simulations of the solar photosphere, and employing simple plate deformation geometries, based on Zernike polynomials: focus (power) and spherical aberrations. The real figure errors may deviate from these simple patterns, but the effect on the measurement quality will be the same. Figure 6 shows a radial cut of the spherical-aberration-like air-gap variation used here. Coating experiments on a full-size prototype Etalon plate have shown that the resulting figure errors occur indeed on very large spatial scales, justifying the simulations made here. The zero line in Figure 6 is chosen such that the air gap averaged across the Etalon aperture has its nominal value; therefore the gap error changes sign. The figure errors of two Etalons cause a mismatch between the transmission peaks across the field of view. This leads to a reduction in total transmission, as shown in Figures 7 and 8. Figure 7 is a gray-scale representation of the full field of view, shown for an air-gap error of 5 nm RMS. Figure 8 show the same result, but as cross-section through the center of the field of view. The effects of the figure errors are imaged to the VTF focal plane; therefore, the horizontal axis is scaled in arcsec. The 250 mm clear aperture of the Etalons translates into a field of view with a diameter of 60 arcsec.

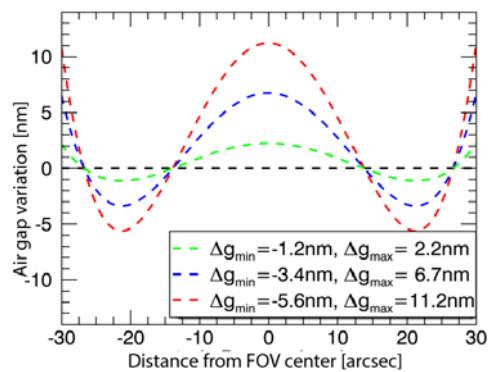

Figure 6. Radial cut across one of the VTF Etalons showing the spherical-type air-gap variations used in this study. Green: 1 nm RMS; blue: 3 nm RMS; red: 5 nm RMS. The horizontal axis is scaled in arcsec in the focal plane; 0 denotes the center of the FOV.

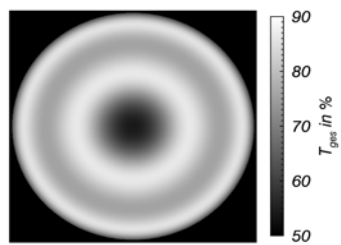

Figure 7: Transmission change caused by large-scale air gap variations of 5 nm RMS[11].

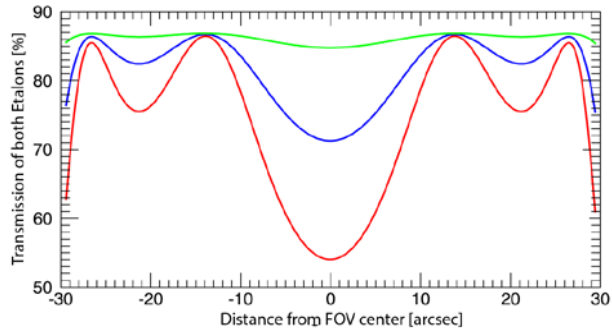

Figure 8: Combined transmission variation for two Etalons, for large-scale cavity errors with RMS amplitudes of 1 nm (green), 3 nm (blue) and 5 nm (red). The slope of the air-gap variation is the same as in Figures 6 and 7.

In addition to the transmission variation shown above, cavity errors also introduce a wavelength shift of a spectral line across the field of view. Figure 9 shows the line shifts, expressed in Doppler velocities that are caused by a 5 nm RMS air-gap error, corresponding to the red line in Figure 8. These large-scale line shifts are a fixed pattern and are easily calibrated in the science data. From the Etalon development program, we conclude that cavity errors of 3 nm RMS or less will be achieved, reducing the effect shown in Figure 9 by a factor of 2, resulting in a large-scale wavelength shift of about 4 pm RMS. At 600 nm, the spectral resolution of the VTF is 6 pm, i.e., the large-scale line shift is little more than one wavelength scan step.

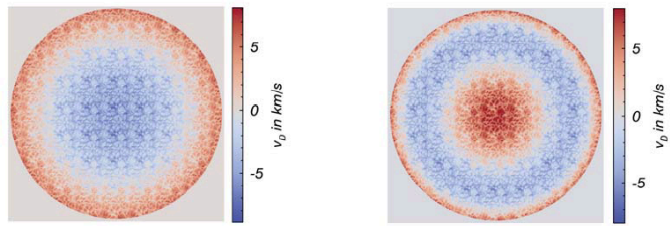

Figure 9: Line shifts caused by a large-scale air-gap variation of 5 nm RMS, expressed in Doppler velocity. At a wavelength of 600 nm, a Doppler shift of 1 km/s is equivalent to a line shift of 2 pm. The small-scale pattern seen in the figure is the (simulated) solar granulation pattern[17].

## 5. SIMULATED SPECTROPOLARIMETRIC OBSERVATIONS

### 5.1 Spectropolarimetry with different Etalon configurations

For the line synthesis, we used a HSRA atmosphere and a simple magnetic field configuration (constant angle, constant field strength). The simulation of the VTF includes the plate finesse, as specified for the VTF, as well as the contribution from the different beam angles of the telecentric configuration with its f:200 beam. The exact location of the etalons with respect to the focal plane is also taken into account. Firstly, we compared simulated spectropolarimetric observations of the Fe 630.25 line using the 1-Etalon and the 2-Etalon setup. Secondly, we investigated simulated observations of the Ca 854.2 nm line using the 1-Etalon setup.

**VTF Simulation**

Stokes profiles for the 630 nm lines were computed for magnetic field strengths between 10 and 1000 Gauss, and inclination angles between 0 (vertical) and 90 degrees. The Stokes profiles were modulated using an "ideal" modulator (see Sect. 3.3). Photon noise with an effective S/N of 675 was added to the modulated intensities $I_1$, to $I_4$. "Effective" refers to the S/N after beam combination, thus the S/N for an individual beam was $675/\sqrt{2} = 475$. For the simulations, it was sufficient to consider only one beam of the polarimeter and then multiply the result by two. The modulated intensities were convolved with the VTF (Etalon + prefilter) transmission profiles. The resulting intensity profiles were then demodulated into Stokes I, Q, U and V using the inverse modulation matrix.

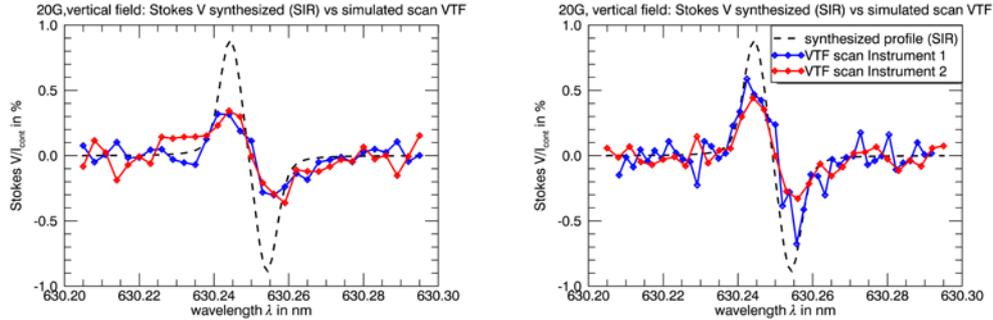

Figure 10: Comparison of synthetic Stokes V profiles of the 630.25 nm line with simulated and observed measurements for different instrument configurations. Left panel: 1-Etalon configuration with 0.2 nm and 0.3 nm prefilters (red and blue line); right panel: 2-Etalon (red) and 3-Etalon (blue) configurations.

**Spectral inversions**

To determine the magnetic field sensitivity, spectral inversions of the simulated observations were performed, and the result of the inversions was compared to the simulated input spectra. SIR was used for the Fe lines, and NICOLE was used for the infrared Calcium line. For each simulated ("observed") line profile, the inversion was repeated 50 times (with slightly varying convergence conditions, noise level) and the mean value of the inferred magnetic field strength was taken. Figure 10 shows a comparison between the simulated (input) spectra, shown as dashed line, and the "observed" ones, using the VTF simulator described in Sect. 3, and with 0.3% photon noise added. The left panel is for a single-Etalon configuration with 0.2 and 0.3 nm prefilters, the right panel corresponds to configurations with two and three Etalons. The input magnetic field strength of 20 G is well retrieved in all cases.

The longitudinal sensitivity was measured for an inclination angle to the vertical of 2° and with magnetic field strengths ranging from 20 G to 100 G for the Fe 630 nm lines, and between 100 and 400 G for the 854.2 nm line. The transversal sensitivity was tested for an inclination of 88° and magnetic field strengths between 80 and 220 G and 400 and 800 G and for the Fe and Ca lines, respectively. The results are shown in Fig. 11 for the Fe 630.25 nm line and in Fig. 12 for the Ca 854 nm line. For the Ca II 854.2 nm line, we used the same VTF parameters, i.e., full spatial resolution, and 8 accumulations with a total integration time of 200 ms. The photon flux in that strong line is lower, and the magnetic sensitivity is much smaller, compared to the 630.25 line, thus the minimum field strength that can be retrieved is higher than for 630.25 nm. Figure 13 therefore shows different magnetic field intervals compared to Figure 12.

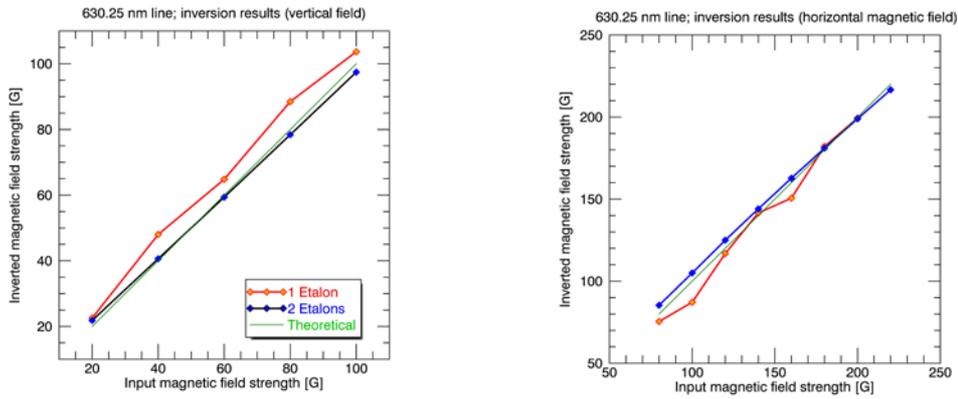

Figure 11: Comparison of the magnetic field sensitivity of the 1-Etalon and the 2-Etalon configurations, using simulated observations of the 630.25 nm line. Red : 1 Etalon; blue:  2 Etalons; green:  nominal slope. Left panel: vertical field component. right panel: horizontal field. component.

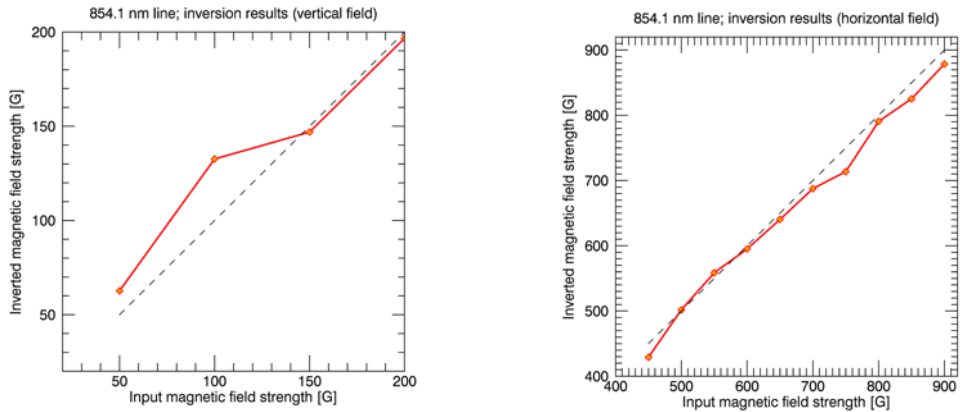

Figure 12: Magnetic field sensitivity for the Ca 854 nm line. Left panel: vertical field component; right panel: horizontal component. The red line shows the relation between input (horizontal axis) and inverted field strength (vertical axis). The dashed line indicates the nominal slope.

## 5.2 Sensitivity of Doppler shift measurements

Figure 13 shows Doppler maps based on an  MHD simulation of solar granulation [11,18]. The simulation box has an area of 5100 km x 5100 km on the Sun (7" x 7" in the sky). To compute the Doppler map that is shown in the top right panel, we convolved the simulated line profile with the spectral resolution of the VTV (6 pm at 630 nm) and employed a spectral sampling of 3 pm. The top right panel shows the Doppler map derived from the simulation. The bottom left map includes 15% of parasitic light (a worst case scenario for a configuration with only one Etalon) and 0.5% of photon noise. The bottom right panel shows the residual velocities, the color table spans a range of ± 0.2 km/s. the RMS-value is 75 m/s. The residual velocities result from the photon noise, i.e., the results are the same for the standard configuration of the VTF with 2 Etalons, which has virtually no parasitic light contribution.

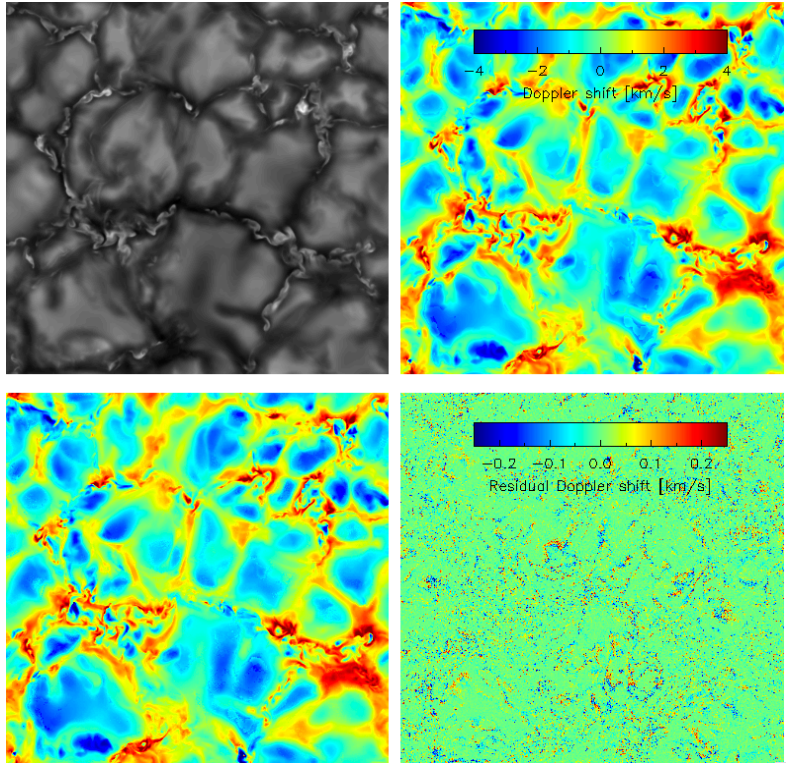

Figure 3: Top left: MHD simulation of solar granulation [11, 18] showing an area of 5100 km x 5100 km on the Sun; top right: Doppler map from simulated data; bottom left: Doppler map, including 15% of parasitic light and 0.5% photon noise; bottom right: residual velocities. Note the different velocity ranges.

## 6. CONCLUSIONS

We investigate the influence of large-scale cavity errors of the Etalons on the instrument performance and conclude that cavity errors of 3 nm RMS are necessary and sufficient to meet the DKIST science requirements for the VTF. These cavity errors cause transmission changes across the field of view, and they lead to systematic wavelength shifts. We show that the transmission changes are of the order of 10%. This value can easily be calibrated. The large-scale line shift is around 4 pm. At 600 nm, the spectral resolution of the VTF is 6 pm, i.e., the large-scale line shift is little more than one wavelength scan step. For a 2-Etalon configuration, the wavelength scan range is at least 0.3 nm, i.e., Doppler shifts up to 150 km/s could be detected.

A polarimetric sensitivity of 20 G for vertical magnetic field and for 150 G for horizontal magnetic field for the Fe 630.25 nm line (g=2.5) is required for the VTF. These numbers were verified by our simulations for instrument configurations with 2 Etalons and also for a single-etalon setup. The latter one requires much narrower prefilters to isolate a single interference order. At the Ca 854.1 nm line, the polarimetric sensitivity is lower, but again within the expected range for both etalon configurations.

A configuration with one high-resolution Etalon is discussed as a first-light option, to meet the overall schedule. A single-Etalon solution would require the usage of ultra-narrowband prefilters. This would reduce the wavelength scan range around the center of each spectral line, and such a solution would also increase the out-of-band contribution of parasitic light considerably to a total amount of about 8.5%. The parasitic light is wavelength-independent and unpolarized. The simulations show that spectropolarimetry both with photospheric and with chromospheric lines is doable with the required sensitivity. The precision of Doppler measurements also meet the science requirements. Due to the very narrow prefilters, Doppler measurements are limited to a range of ± 20 km/s. Monochromatic intensity imaging in the cores of very strong spectral lines would strongly suffer from the level of parasitic light. The throughput of a VTF with a single Etalon would be reduced by about 10%, with corresponding impact to either the SNR or the measurement cadence.